\documentclass[twocolumn,showpacs,preprintnumbers,amsmath,amssymb,floatfix]{revtex4}

\usepackage{graphicx}
\usepackage{dcolumn}
\usepackage{bm}

\newcommand{\beq}{\begin{equation}}
\newcommand{\eeq}{\end{equation}}
\newcommand{\bey}{\begin{eqnarray}}
\newcommand{\eey}{\end{eqnarray}}

\begin{document}

\title{ A new proposal for Galactic dark matter: Effect of f(T)
gravity}

\author{Farook Rahaman}
\email{rahaman@iucaa.ernet.in} \affiliation{Department of
Mathematics, Jadavpur University, Kolkata 700 032, West Bengal,
India}
\author{Ritabrata Biswas}
\email{ritabrata@physics.iisc.ernet.in
 } \affiliation{
Department of Physics, IISc, Bangalore, India}

\author{Hafiza Ismat Fatima}
\email{hafizaismatfatima@yahoo.com} \affiliation{Department of
Mathematics, Govt. Degree college (W) Warburton, Nankana Sahib,
Pakistan}

\author{Nasarul Islam}
\email{nasaiitk@gmail.com}\affiliation{Department of Mathematics,
Danga High Madrasah, Kolkata 700 103, West Bengal, India}
\date{\today}

\begin{abstract}
\noindent It is still a challenging problem to the theoretical
physicists to know the exact nature of the galactic
 dark matter which causes the galactic rotational velocity to be more or less a
 constant. We have proposed   the dark matter as an effect of f(T)
 gravity. Assuming the flat rotation curves as input we have shown
 that f(T) gravity can explain galactic dynamics. Here, we do not
 have to introduce dark matter. Spacetime metric inspired by f(T)
 gravity describes the region up to which the tangential velocity
 of the test particle is constant. This inherent property appears
 to be enough to produce stable circular orbits as well as
 attractive gravity.

\end{abstract}


\maketitle

 \section{Introduction}
\noindent The sources that are responsible for the rotation curves
of neutral hydrogen clouds in the outer regions of  galaxy remain an
active area of research in recent time. The observed luminous matter
content of the galaxies does not follow this behavior. In other
words, either Newtonian gravity and general theory of gravity fail
to explain this phenomena or there exists some non luminous matter
hidden in the galaxy. About 80 years ago, Zwicky \cite{jO32}
confirmed the existence of non luminous matter, so called dark
matter, by comparing the virial mass and luminous mass of the
galactic clusters. The nature of this non luminous matter contained
in the galaxy is still unknown as this matter does not follow
particle standard model. In spite of several analytic models are
proposed time to time. For a summary of some alternative theories
such as  noncommutative-geometry background, scalar-tensor theory,
 brane-world models, see the references \cite{jO33}-\cite{jO41}.
Some authors have considered  quintessential matter, perfect fluid
type and global monopoles as a candidate for galactic halo matter
\cite{jO42}-\cite{jO44}.

Teleparallel equivalent of general relativity (TEGR) received a
great attention as alternative theory of gravity. In this new
theory, it is assumed that the manifold contains, in addition to
curvature, a quantity called torsion. This theory is known as
$f(T)$ theory. In analogous to the $f(R)$ theory, $f(T)$ theory is
constructed with a generalized Lagrangian that deviates from
Einstein gravity by a function $f(T)$, where $T$ is the so-called
torsion scalar \cite{jO45}-\cite{jO46}. Recently,   some authors
\cite{jO47} show that the $f(T)$ gravity theory is extensively
used to explain the accelerated state of the Universe without
introducing the dark energy.

Deliduman and Yapiskan \cite{jO49} have observed the constraint on
the neutron solution in $f(T)$ theory of gravity and concluded
that the solution is possible only for a linear function of
torsion. Cai et al. \cite{jO50} have  discussed the matter bounce
cosmology with scalar and tensor modes of cosmological
perturbations. B\"{o}hmer et al. \cite{jO48} have investigated the
existence of relativistic stars taking different approaches in
this theory and have constructed static perfect fluid solutions.

In this paper, we propose the source of the galactic halo
 as an effect of $f(T)$ gravity
in addition to the normal matter as luminous matter. All the astrophysical observations made in last one and half decades
 require the modification of either side of the Einstein gravity  for their theoretical explanations. Galactic halo is popularly
  explained by the effect of dark matter . However, as yet no
convincing argument about the origin of the angular momentum has
been suggested. Hoyle \cite{Hoyle1} first proposed gravitational
instability, arising from gravitational coupling with the
surrounding matter (tidal interactions), as a possible explanation
for galactic rotation. Alternatively, peoples have proposed
\cite{vonWiezsacker1, Gamow1} that the origin of galaxy rotation
could be due to primordial turbulence/vorticity.
However, vortical modes decay with time and, therefore, a
significant vorticity at the time of galaxy formation would
typically require an unrealistic magnitude of vorticity in the
early universe. At present, it is widely believed that the
hierarchical clustering of cold dark matter \cite{Blumenthal1} is
the origin of structures in the universe. Consequently, the
angular momentum of dark matter halos and eventually the rotation
of galaxies is thought to be produced by gravitational tidal
torque \cite{Barnes1}. But inclusion of Dark matter means the
modification of the energy momentum tensor in Einstein equation.
Now in spite of doing that we are free to modify the geometry side
of the Einstein equation to support the same observation. It means
we can modify our gravity. Here we have chosen the TEGR as a
modification of gravity. Now  our motive is to replace the dark
matter by modified gravity.  It is argued that by using modified
gravity, one can find
 gravitational analog of the
Lorentz force equation, which is an equation written in the
underlying Weitzenbock spacetime \cite{ANDRADE1}. According to
this approach, the trajectory of the particle is described in the
very same way the Lorentz force describes the trajectory of a
charged particle in the presence of an electromagnetic field, with
torsion playing the role of force. These all uses made us more
interested to find a way to try to interpret the galactic halo as
a result of TEGR rather than dark matter.

The outline of the paper is as follows: In the next section, we
discuss the galactic rotation curves using a spherical gravitational
field. Section III provides the field equations in $f(T)$ theory and
their solutions. Section {IV discusses the results of the solutions
for $\frac{ d f}{dT}=1$ and $ \frac{ d f}{dT}\neq 1$ and gives several observations
following the obtained solutions. The last section V contains
concluding remarks about the proposal.

\section{\textbf {Galactic rotation curves}}

The observation of the motion of neutral hydrogen clouds are
believed as evidence for the existence of some kind of exotic
matter in the galactic halo apart from luminous matter in the
galaxies. These hydrogen clouds in the outer region of galaxies
are treated as test particles moving in circular orbits due to
gravitational effects of the combined matters ( luminous matter as
normal matter and non luminous matter as dark matter ) contained
in the galaxies. The gravitational field inside the halo is
characterized by line element

\begin{equation}\label{line1}
ds^2=-e^{\nu(r)}dt^2 + e^{\lambda(r)}dr^2+r^2 d\Omega^2,
\end{equation}
where $d\Omega^2=d\theta^2+\text{sin}^2\theta\, d\phi^2$. We are
using here geometrized units in which $G=c=1$.

To study the  tangential velocity of  such  circular orbits, we will
derive geodesic equation from the    Lagrangian for a test
particle travelling on the spacetime (\ref{line1})  which  is given by
\begin{equation}\label{E:Lagrangian2}
2\mathcal{L}=-e^{\nu(r)}\dot{t}^{2}+e^{\lambda(r)}\dot{r}^{2}+r^{2}\dot{\Omega}^{2},
\end{equation}
where as usual
$d \dot{\Omega}^2=d \dot{\theta}^2+\text{sin}^2\theta\,d\dot{\phi}^2$.
The overdot denotes differentiation with respect to affine
parameter $s$.

From  (\ref{E:Lagrangian2}), the geodesic  equation  for material particle can be
written as
\begin{equation}\label{E:motion}
\dot{r}^{2}+V(r)=0
\end{equation}
now yields the potential
\begin{equation}\label{E:V2}
V(r)=-e^{-\lambda(r)}\left(e^{-\nu(r)}E^{2}-\frac{L^{2}}{r^{2}}-1\right).
\end{equation}

Here the conserved quantities $E$ and $L$, the energy and total
momentum, respectively are given by  $E=-e^{\nu(r)}\dot{t}$,
$L_{\theta}=r^2\dot{\theta}$, and
$L_{\phi}=r^{2}\text{sin}^{2}\theta\,\dot{\phi}$. So the square of
the total angular momentum is $L^{2}=
{L_{\theta}}^{2}+(L_{\phi}/\sin\theta) ^{2}$.

Following \cite{jO33,jO34}, one can find the tangential velocity
of the test particle,
\begin{equation}\label{E:tangential1}
(v^{\phi})^2=\frac{e^{\nu}L^2}{r^2E^2}=\frac{1}{2}\,r\,\nu'.
\end{equation}
This expression can be integrated to yield the metric coefficient
at the distances where the circular velocity remains constant with radius
\begin{equation}\label{E:tangential2}
e^{\nu}=B_0r^l,
\end{equation}
where $B_0$ is an integration constant and $l=2(v^{\phi})^2$.

Equivalently, the line element (\ref{line1}) in regions with a flat rotational curve
\begin{equation}\label{E:line2}
ds^2=-B_0r^ldt^2+e^{\lambda(r)}dr^2+r^2d\Omega^2.
\end{equation}

It seems to me  the source of galactic matter  as an effect of
$f(T)$ gravity in addition to the normal matter as luminous matter.
Therefore, it is interesting to find spacetime metric of the
region of constant tangent velocity.

\section{  \textbf{   Field Equations in  $f(T)$ theory :} }
Similar to $f(R)$ modified theory of gravity, the action of $f(T)$
theory is taken as \cite{jO45}-\cite{jO48}
\begin{equation}\label{euation8}
S[e^i_\mu , \phi_A ] = \int d^4x \left[ \frac{1}{16 \pi} f(T) +
L_{matter} (\phi_A)  \right]
\end{equation}
Here, $e^i_\mu$ are the tetrad by which we can define any metric
as $g_{\mu \nu} =  \eta_{ij} e^i_\mu e^j_\nu$, where $ \eta_{ij} =
diag ( -1,1,1,1 )$ and $e_i^\mu e^i_\nu = \delta_\nu^\mu $, $ e=
\sqrt{-g} = det(e^i_\mu)$.\\
We used the units $G=c=1$. Here, $\phi_A$  are matter fields and
$f(T)$ is an arbitrary analytic function of the torsion scalar
$T$. The torsion scalar is built from torsion and contorsion as
\begin{equation}\label{euation9}
T = S_\sigma^{ \mu\nu} T^\sigma_{ \mu\nu}
\end{equation}
where,
\begin{equation}\label{euation10}
T^\sigma_{ \mu\nu} =   \Gamma^\sigma_{ \mu\nu}-\Gamma^\sigma_{
\nu\mu} = e_i^\sigma \left( \partial_\mu e^i_\nu - \partial_\nu
e^i_\mu \right)
\end{equation}
\begin{equation}\label{euation11}
K_\sigma^{ \mu\nu} =  -\frac{1}{2} \left( T^{ \mu\nu}_{~~~\sigma}-
T_{~~~\sigma}^{ \nu\mu}-T_{\sigma}^{ \mu\nu}  \right)
\end{equation}
are torsion and contorsion respectively with new defined tensor
components
\begin{equation}\label{euation12}
S_\sigma^{ \mu\nu} =  \frac{1}{2} \left( K^{ \mu\nu}_{~~~\sigma}+
\delta_\sigma^\mu T ^{ \beta \nu}_{~~~\beta}-\delta_\sigma^\nu T
^{ \beta \mu }_{~~~\beta} \right)
\end{equation}
Varying the action (\ref{euation8}) with respect to the tetrads, one
can obtain the field equations of $f(T)$ gravity as
\begin{equation}\label{euation13}
S_i^{ \mu \nu}f_{TT} \partial_\mu T + e^{-1} \partial_\mu ( e
S_i^{ \mu \nu}) f_T -T^\sigma_{ \mu i}S_\sigma^{ \nu \mu} f_T +
\frac{1}{4} e_i^\nu f  = 4 \pi \Upsilon_i^\nu
\end{equation}
where,
\[ S_i^{ \mu \nu} =e_i^\sigma S_\sigma^{ \mu \nu}, ~~f_T =
\frac{\partial f}{\partial T}~~~ and  ~~~f_{TT} = \frac{\partial^2
f}{\partial T^2}~~\] and $\Upsilon_i^\nu$ , the energy momentum
tensor of the anisotropic fluid is given by
\[
  \Upsilon_\nu^\mu=  ( \rho + p_r)u^{\mu}u_{\nu} - p_r g^{\mu}_{\nu}+
            (p_t -p_r )\eta^{\mu}\eta_{\nu}
\]
with $$ u^{\mu}u_{\mu} = - \eta^{\mu}\eta_{\mu} = 1. $$

For our metric (\ref{line1}), we define the tetrad matrix as
\begin{equation}\label{euation14}
e^i_{ \mu} = ~diag [ \sqrt{B_0 r^l},~ e^{\frac{\lambda}{2}},~ r, ~r
\sin\theta ] .
\end{equation}
Now, we compute the torsion scalar as
\begin{equation}\label{euation15}
T(r) = - \frac{2 e^{-\lambda}}{r} \left( \frac{l}{r} + \frac{1}{r}
\right)
\end{equation}
Inserting this and the components of the tensors $ S_i^{ \mu \nu}$
 and $T_i^{ \mu \nu}$  in (\ref{euation13}) yields,
\begin{equation}\label{euation16}
4 \pi \rho = \frac{ e^{-\lambda}}{r} T^{\prime} f_{TT} - \left [
\frac{T}{2} + \frac{ 1}{2r^2} + \frac{ e^{-\lambda}}{2r} \left(
\frac{l}{r} + \lambda^\prime \right) \right]f_T + \frac{f}{4}
\end{equation}
\begin{equation}\label{euation17}
4 \pi p_r =   \left [ \frac{T}{2} + \frac{ 1}{2r^2} \right]f_T -
\frac{f}{4}
\end{equation}
$
4 \pi p_t =  -\frac{e^{-\lambda}}{2} \left (\frac{l}{2r} +
\frac{1}{r} \right)T^{\prime} f_{TT} + \frac{T }{4} f_T
\\
\\~~~~~~~~~~~-\frac{e^{-\lambda}}{2} \left [\frac{-l}{2r^2} + \left(
\frac{l}{4r} +\frac{1}{2r} \right) \left( \frac{l}{r}
-\lambda^\prime \right) \right]f_T -\frac{f}{4}$
\begin{equation}\label{euation18} \end{equation}

According to  B\"{o}hmer et al \cite{jO48}, one of the possibility
of recovering general relativity result is $f_{TT} = 0$, though
its general relativity form has no meaning in the present context.

\section{Solutions and their consequences}

 Now
we are seeking solutions using some specific forms of f(T).
\\
\\
\subsection{$f(T) =  T$  }

Let us seek the influence of f(T) gravity by assuming  $f_{T} =
1$. This assumption immediately yields
\begin{equation}\label{euation19}
f(T) = T
\end{equation}
which immediately follows $f_{TT} = 0$. For normal matter, we have
\begin{equation}\label{euation20}
p_r = m \rho
\end{equation}
where, $'m'$ is an equation of state parameter.\\

 Now, using equation (\ref{euation19}), solving $e^{-\lambda(r)}$ from (\ref{euation20}), (\ref{euation16}) and (\ref{euation17}) we have,
\begin{equation}\label{euation21}
e^{-\lambda(r)}=  \left[\frac{1+m}{m+l+1}\right]
+Dr^{-\left[\frac{m+l+1}{m}\right]}
\end{equation}
where, $'D'$ is an integration constant.
Then, from equation (15) we can obtain
\begin{equation}\label{euation22}
T(r) = - \left[\frac{2(1+m)(1+l)}{(m+l+1)}\right]r^{-2}
-2D(l+1)r^{-\left[\frac{3m+l+1}{m}\right]},
\end{equation}
\begin{equation} 4 \pi \rho(r)=-\left[\frac{ D(l+1)}{2m}
\right]r^{-\left[\frac{3m+l+1}{m}\right]}
+\left[\frac{l}{2(m+l+1)}\right] r^{-2}   \label{euation23}
\end{equation} ~and~
\\
\\
 $ 4 \pi p_t= \left[\frac{ D(l+2)(m+l+1)}{16 m (1+l)
}\right]r^{-\left[\frac{3m+l+1}{m}\right]}\\
\\~~~~~~~~~~~~~~~~~~~ - \frac{l^2}{8 r^2} \left[\left(\frac{1+m}{m+l+1}\right)
+Dr^{-\left[\frac{m+l+1}{m}\right]} \right]$
\begin{equation} \label{euation24} \end{equation}

 \textbf{Comments on the solutions :}
 \\
 \\
\textbf{(a)} Notice that this type of spacetime definitely cannot
be asymptotically flat neither can it have the form of a spacetime
due to a centrally symmetric black hole. Since this space
describes the region up to which tangential velocity is constant,
so it has to be joined with exterior region with other types of
spacetime, may be with Schwarzschild spacetime. Thus, in
principle, one can obtain the value of $D$ with the junction
conditions. However, it is yet to discover the galactic boundary.

\textbf{(b)} The remarkable effect due to $f(T)$ gravity is the presence of first term in the
expression of energy density .
According to Newtonian theory which is indistinguishable from
general relativity in very weak field, to maintain circular
orbits of the neutral hydrogen clouds with constant velocity, the
centrifugal acceleration $\frac{(v^{\phi})^2}{r}$ should be equal
to the gravitational attraction $\frac{G M(r)}{r^2}$ of the total
mass $M(r)$ contained within radius $r$. This yields the Newtonian
mass $M(r)$ is increasing linearly with $r$. In other words,
$\rho_{Newton} \propto \frac{1}{r^2}$. The first term in the
expression of energy density due to the effect of $f(T)$ gravity.
This contribution will vanish if $D$ vanishes.

\textbf{(c)} Now, we try to recognize  $D$. Our metric (\ref{E:line2}) will take
the following in the limit  $l \rightarrow 0$ as $ ds^2=-dt^2
     +\frac{1}{1 + D r^{-\frac{m+1}{m}}} dr^2+r^2d\Omega^2
 $. The standard FLRW metric is $ ds^2=-dt^2
     +\frac{1}{1 - kr^2 }dr^2+r^2d\Omega^2
 $ . If one considers, $ m = -\frac{1}{3}$, then, one can lead
 identification that $D = -k$, in other words,
 $D$ is recognized as the spatial curvature (with a
 negative sign). Thus, one may assume our obtained spacetime metric of the galactic halo
 is induced by the space-time embedded in a static FLRW
 metric. It is known that spatially curved FLRW universe would be
 static and stable if the universe is supported by the fluid with
 equation of state $ \rho = -3p$. The dynamicity of FLRW metric is
 absent here, as we are working on a local problem like flat rotation
 curve which fixes the scale factor $R(t)$ to be constant. The fluid
 following
 equation of state $ \rho = -3p$ is not exotic as it follows null
 and weak energy condition. Hence, we don't need to consider dark
 matter as exotic matter. It is nothing but an effect of $f(T)$
 theory.
Now, we calculate Ricci scalar ($R_c$) for the derived spacetime
which is given by,
\begin{widetext}
\begin{eqnarray}
R_c=\frac{D\left[\left(\frac{m+l+1}{m}\right)^{2}(4+l)\right]-\left[\left(\frac{m+l+1}{m}\right)D+\left(\frac{m+1}{m}\right)
r^{\left(\frac{m+l+1}{m}\right)}\right](l^{2}+2l+4)
+4\left(\frac{m+l+1}{m}\right)r^{\left(\frac{m+l+1}{m}\right)}}{2\left[\left(\frac{m+l+1}{m}\right)r^{2+\left(\frac{m+l+1}{m}\right)}\right]}
\end{eqnarray}
\end{widetext}
As $l\rightarrow 0$  and $m=-\frac{1}{3}$,  $R_c =-6D$, once again
suggesting that $D$ is the spatial curvature.

 However, in general $l\neq 0$ and hence the above
 interpretation is not universally truth.\\
\\
\textbf{(d)} If we write $e^\lambda$ in the standard Schwarzschild
form
\[
e^\lambda=\left[1-\frac{2{\cal M}(r)}{r}\right]^{-1},
\] then our model reveals that the proper radial length is larger
than the Euclidean length because $r > 2{\cal M}(r)$. This is a very
crucial condition for a valid metric as this condition keeps safe
the signature. For example, if we put $D=0 $ in our metric
function (21),  we get,
\[ e^\lambda = \frac{1+m+l}{1+m} = 1 + \frac{l}{1+m}  ~ > 1\]
This implies  $D =0$ is one of the possible values which fulfils the
essential requirement.
\\
\\
\textbf{(e)}    ~After performing the radial rescaling \[ r=
\sqrt{\frac{1+m}{1+m+l}} R,\] one can rewrite the metric (7) for
the solution (22) with $D=0$ as
\[ ds^2 =
 - B_0^\prime R^l dt^2 + dR^2 +\left[ \frac{1+m}{1+m+l} \right]R^2 ( d \theta^2
 + \sin^\theta d\phi^2 )\]
where, $B_0^\prime = B_0 \left(\sqrt{\frac{1+m}{1+m+l}}\right)^l$.
\\
\\
This shows an angle of surplus ( analogous  to the angle of
deficit ) in the surface area given by
\[S_1 = 4 \pi R^2 \left[ \frac{1+m}{1+m+l} \right] = \left[ \frac{1+m}{1+m+2(v^{\phi})^2} \right]  4 \pi R^2 \]
One can observe that if the rotational velocity of the probe
particles wee photons so that $v^{\phi}\rightarrow 1$, the surface
would remain finite, however reduced to the factor $\left[
\frac{1+m}{3+m} \right] $ of the spherical surface area $S_2 = 4
\pi R^2$. However, for  a typical galaxy, the rotational velocity
is $v^{\phi} \approx 10^{-3}$ and in this case the difference of
two surface areas
\[  S_2-S_1 =4 \pi R^2 \left[ \frac{2(v^{\phi})^2}{1+m+2(v^{\phi})^2} \right]   \]
One can note that for increasing the value of equation of state
parameter, this difference decreases. Thus rate of the difference
in units of flat surface area grows as $\approx 10^{-6}$.
Definitely, nowadays, this measurement is sufficient to examine
the actual behavior of the motion of stars in a galaxy.
\\
\\
\textbf{(f)} It can be verified that $ \rho > 0,~\rho+p_r >0 ~,
\rho+p_r+2p_t >0$. This means halo contains non exotic matter as
matter satisfies all energy conditions. Therefore, we expect an
attractive halo. Following  Lynden-Bell et al's \cite{jO51}
prescription we ratify it by calculating the total gravitational
energy $E_G$ which is given by
\[ E_{G}=M-E_{M}=4\pi\int_{r_{1}}^{r_{2}}[1-\sqrt{e^{\lambda(r)}}]\rho r^{2}dr \]
Note that, since $\rho >0$, $1-\sqrt{e^{\lambda(r)}} <0$ and $r_2>
r_1$,  therefore, $E_G <0$, showing that gravity in the halo is
indeed attractive.\\
Also, one can confirm that the effect is attractive by studying the geodesic
equation
\[
\frac{d^{2}x^{\alpha}}{d\tau^{2}}+\Gamma_{\alpha}^{\mu\gamma}\frac{dx^{\mu}%
}{d\tau}\frac{dx^{\gamma}}{d\tau}=0
 \]
for a test particle that has placed at some radius $r_0$. This
yields the radial equation
\[  \frac{d^2 r} {d\tau^2} = - \frac{1}{2} e^{-\lambda}
    \left[\frac{d }{dr}e^\lambda \left(\frac{dr}{d\tau}
\right)^2 + \frac{d }{dr}e^\nu\left(\frac{dt}{d\tau}
\right)^2\right] \].   As long as $\frac{dr}{d\tau}=0$, we get
\[
  \frac{d^2 r} {d\tau^2} = - \frac{1}{2} e^{-\lambda}
   B_0lr^{l-1}\left(\frac{dt}{d\tau}\right)^2<0.
\]
We conclude that objects are attracted toward the center. This
result agrees with the observational evidence i.e., gravity on the
galactic scale is attractive ( clustering , structure formation
etc ).
\\
\\
\textbf{(g)} ~To check stability of circular orbits, let us assume
a test particle with four velocity
$U^{\alpha}=\frac{dx^{\sigma}}{d\tau}$ moving in the region of
spacetime given in (7). Assuming $\theta=\pi/2$, the equation
$g_{\nu\sigma}U^{\nu}U^{\sigma}=-m_{0}^{2}$ yields
\begin{equation}
\left( \frac{dr}{d\tau}\right)  ^{2}=E^{2}+V(r),
\end{equation}
with
\begin{equation}
V(r)=-\left[E^{2}\left( 1-\frac{r^{-l}e^{-\lambda} }{B_{0}}\right)
+e^{-\lambda}\left( 1+\frac{L^{2}}{r^{2}}\right) \right].
 \end{equation}
where, $ E=\frac{U_{0}}{m_{0}}\quad $ and $ L=\frac{U_{3}}{m_{0}}$
are two conserved quantities, namely relativistic energy (E) and
angular momentum (L) per unit rest mass of the test particle
respectively. If the circular orbits are defined by $r=R$, then
$\frac{dR_0}{d\tau}=0$ and, additionally,
$\frac{dV}{dr}\mid_{r=R}=0$. Above two conditions result
\begin{equation}
L=\pm\sqrt{\frac{l}{2-l}}R
\end{equation}
and, using $L$ in $V(R)=-E^{2}$, we get
\begin{equation}
E=\pm\sqrt{\frac{2B_{0}}{2-l}}R^{\frac{l}{2}}.
\end{equation}
The orbits will be stable if $\frac{d^{2}V}{dr^{2}}\mid_{r=R}<0$
and unstable if $\frac{d^{2}V}{dr^{2}}\mid_{r=R}>0$.

By putting the expressions for $L$ and $E$ in
$\frac{d^{2}V}{dr^{2}}\mid_{r=R}$ and then by using the solution
(14), we get,
 \\
 \\
 $ \frac{d^{2}V}{dr^{2}} \mid_{r=R}=-\left[\frac{2l(1+m)}{(1+m+l) R^2}+\frac{ Dl( 3m +l+1)
}{m(2-l)(1+l)} R^{\frac{-(3m+l-1)}{m}}\right] $ \\
\\

One may note that, for $D=0$ as well as for $D\neq0$, $\frac{d^{2}V}{dr^{2}%
}\mid_{r=R}<0$. So the circular orbits are always stable.
\\
\subsection{$f(T) = \alpha T +T_0$}

In the previous section, we have investigated the influence of
f(T) gravity with $f_T = 1$. Now, we will assume the case $f_T
\neq 1$ and consider

\begin{equation}
f(T) = \alpha T +T_0
\end{equation}
where $T_0$ and $\alpha$ are constants.

Again, one can solve for $e^{-\lambda(r)}$ from the field
equations (16) - (18) by using equation(30) as
\begin{equation}
e^{-\lambda(r)}=\frac{1+m}{1+m+l}+D_{2}r^{-\frac{l+m+1}{m}}-\frac{T_{0}(1+m)}{2\alpha
(3m+l+1)}r^{2}
\end{equation}
$D_{2}$ being constant of integration.\\ Accordingly, we have, \\
\\$ T(r)=\frac{T_{0}(1+m)(l+1)}{\alpha
(3m+l+1)}-\frac{2(1+m)(1+l)}{1+m+l}r^{-2}\\~~~~~~~~~~~~~~~~~~~~~~~~~~~~~~~~~~~~~-2D_{2}(l+1)r^{-\frac{l+3m+1}{m}}$
\begin{equation} \end{equation}
\\
\\
\begin{widetext}
\begin{eqnarray}
4\pi \rho=\frac{-\alpha}{4}\left[\frac{T_{0}(1+m)(l+1)}{\alpha
(3m+l+1)}-\frac{2(1+m)(1+l)}{1+m+l}r^{-2}-2D_{2}(l+1)
r^{-\frac{l+3m+1}{m}}\right]-\frac{\alpha r^{-2}}{2}-\frac{l
\alpha (1+m)}{2(l+m+1)}r^{-2}\nonumber
\\
-\frac{\alpha}{2}\left(\frac{l+m(l+1)+1}{m}\right)D_{2}r^{-\frac{3m+l+1}{m}}+\frac{T_{0}}{4}\left(\frac{(l+1)m+2l}{3m+l+1}\right)
\end{eqnarray}
\end{widetext}
\begin{widetext}
\begin{eqnarray}
16\pi p_{r}=\alpha \left[\frac{T_{0}(1+m)(l+1)}{\alpha
(3m+l+1)}-\frac{2(1+m)(1+l)}{1+m+l}r^{-2}-2D_{2}(l+1)r^{-\frac{l+3m+1}{m}}\right]+2\alpha
r^{-2}-T_{0}
\end{eqnarray}
\end{widetext}
\begin{widetext}
\begin{eqnarray}16\pi p_{t}=-T_{0}-\frac{\alpha l^{2}}{2}\left[\left(\frac{1+m}{1+m+l}\right)r^{-2}+D_{2}
r^{-\frac{l+3m+1}{m}}-\frac{T_{0}(1+m)}{2\alpha
(5m+l+1)}\right]\nonumber
\\
+\frac{\alpha
(l+2)}{2}\left[\frac{D_{2}(l+m+1)}{m}r^{-\frac{3m+l+1}{m}}+\frac{T_{0}(1+m)}{\alpha\left(3m+l+1\right)}\right]
\end{eqnarray}
\end{widetext}

 \textbf{Several observations follow from the
solutions:} \\
\\
\\
\begin{figure}
\includegraphics[scale=.3]{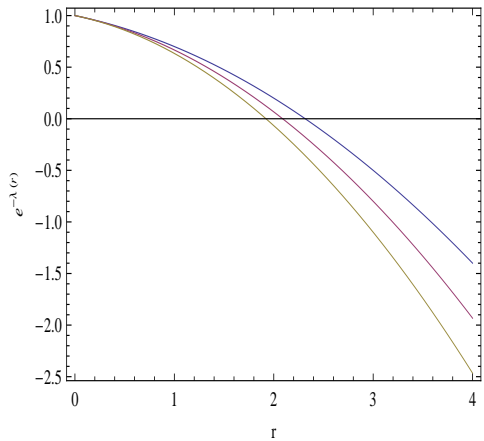}
\caption{The singularity occurs where  $e^{-\lambda}$ cuts r
axis.}
\end{figure}

\textbf{(a)} Point to be noticed is that the solution indicates almost the same kind of space time denoted by the Case $f(T)=T$ solution.\\

\textbf{(b)} We can check in the energy density's expression, the second term in first bracket only the term not containing
 any information from $f(T)$  gravity. Even this guarantees the fact $\rho_{Newton}\propto \frac{1}{r^{2}}$.  Making
 $\alpha=1$ and $D_2=T_0=0$ together, we can go back to the Newtonian solution. \\

\textbf{(c)} While physically interpreting $D_{2}$, we should
check for the metric (\ref{E:line2}) for $l\rightarrow 0$ and
comparing with FLRW metric, we have for $m=-\frac{1}{3}$,
$\frac{T_{0}}{2\alpha}+D_{2}=-\kappa$, $\kappa$ being the scalar
curvature. Interestingly, for  $D_{2}=-\frac{T_{0}}{2\alpha}$, the
spacetime would be flat.
It is to be noted that we are at the Strong energy condition boundary with such an $m$.\\

\textbf{(d)} It is quite obvious that we will not be able to find
exact form of black hole singularity with such a result as trying
to make the $e^{-\lambda(r)}=0$,  will lead us to a transcendental
equation. If we plot $e^{-\lambda(r)}~vs~r$ to find its zeroes,
then we can find some lapse points. Here in figure (1),  one can
see three curves representing  $e^{-\lambda(r)} $ :  the blue one
indicating the curve for $(\alpha=0.1, ~T_{0}=-0.1)$, the maroon
one denotes $(\alpha=0.5, ~T_{0}=-0.3)$ and finally the golden
yellow denotes $(\alpha=1, ~T_{0}=-0.8)$ ($D_{2}$ has been taken
as $-0.2$ in all cases). In all the three  cases,  we can see that
our $g_{rr}$ meets a singularity.

\section{ Conclusion }

In this paper, we have proposed a theoretical possibility of dark
matter as an effect of $f(T)$ theory. Here, we don't have to
consider the exotic matter (i.e. violates the usual energy
condition )  hidden in the galactic halo. Employing certain
restriction on the equation of state parameter of the luminous
matter contained in the galactic region, we have shown flat
rotation curve suggests the background geometry of the universe,
in other
 words, spatial curvature of the universe can be obtained from a local gravitational phenomena.
 It is shown
that the intrinsic property of the $f(T)$ inspired gravity is
able to describe two crucial physical requirements - the stable circular orbits of the test particle out side of the halo
 and  attractive gravity in the halo region.
 The approach used in this
 paper is totally different from other proposals exist in literature. We hope our model inspires the authors to give theoretical
 support of galactic dark matter and
f(T) gravity is a matter of further investigation.\\\\

\pagebreak

{\bf Acknowledgement : }\\\\
RB thanks ISRO grant ``ISRO/RES/2/367/10-11" for providing Research
Associate Fellowship.  FR  gratefully acknowledge support from
IUCAA, Pune, India under Visiting Associateship under which a part
of this work was carried out.  FR is also thankful to PURSE, DST
and UGC, Govt. of India for providing financial support.

\end{document}